# Energy, phonon, and dynamic stability criteria of 2d materials

Oleksandr I. Malyi[1*], Kostiantyn V. Sopiha[2], and Clas Persson[1*]

1 – Centre for Materials Science and Nanotechnology, Department of Physics, University of Oslo, P.O. Box 1048 Blindern, NO-0316 Oslo, Norway
2 – Ångström Solar Center, Solid State Electronics, Department of Engineering Sciences, Uppsala University, Box 534, SE-75121 Uppsala, Sweden

email: oleksandrmalyi@gmail.com (O.I.M), clas.persson@fys.uio.no (C.P)

**Abstract**
First-principles calculations have become a powerful tool to exclude the Edisonian approach in search of novel 2d materials. However, no universal first-principles criteria to examine the realizability of hypothetical 2d materials have been established in the literature yet. Because of this, and since the calculations are always performed in an artificial simulation environment, one can unintentionally study compounds that do not exist in the experiments. Although investigations of physics and chemistry of unrealizable materials can provide some fundamental knowledge, the discussion of their applications can mislead experimentalists for years and increase the gap between experimental and theoretical research. By analyzing energy convex hull, phonon spectra, and structure evolution during ab initio molecular dynamics simulations for a range of synthesized and recently proposed 2d materials, we construct energy, phonon, and dynamic stability filters which need to be satisfied before proposing novel 2d compounds. We demonstrate the power of the suggested filters for several selected 2d systems, revealing that some of them cannot be ever realized experimentally.

**Introduction:** With the ever-growing interest in 2d materials, the search for new materials and understanding their basic properties via first-principles calculations have become a paramount topic in physics/chemistry of low-dimensional systems. In principle, the theoretical studies can exclude a large portion of the Edisonian approach guiding experimental work towards new materials synthesis for new advanced applications. In particular, recent state-of-art high-throughput screening[1-4] have provided basic information on hundreds on new 2d materials that will be synthesized with time. However, in general, the calculations are always performed in an artificial simulation environment. From one side, it allows exploring fundamental materials properties excluding external factors. From the other side, one can study systems which can never be realized experimentally due to the strong tendency for decomposition/reconstruction or a large amount of energy needed to realize the free-standing layers. Although investigations of unstable systems can provide some fundamental physics/chemistry and understanding of materials properties, potential discussion of their applications can mislead experimentalists for years.[5-6] Even though some thermodynamically unstable free-standing layers can be stabilized via epitaxial growth on the selective substrates[7], such systems cannot be modeled as free-standing layers since this approximation would neglect interaction at the formed interface which potentially determines system properties. Because of this, there is a clear gap between many theoretical studies and their experimental verifications. Moreover, there is still no universal framework to understand stability of the proposed 2d materials. The majority of first-principles studies still consider negative formation enthalpy (heat), negative cohesive energy, or absence of negative vibrational modes to be the main parameters determining the stability of 2d materials. Although these parameters are critical, in most cases, they are not sufficient to ensure that the materials can ever be obtained experimentally. This uncertainty has led to the situation where several polymorphs of the same material are often studied independently, thus providing completely different conclusions. Considering a current interest in search of novel functional materials, it is vital to establish a generalized framework (*i.e.,* the "*game rules*") to unambiguously explore 2d materials stability at the first-principles level. Motivated by this, we explain herein some critical factors that must be considered during analysis of the stability of hypothetical 2d materials and several instances where these factors play a major role.

**Methods:** All calculations are carried out using the Vienna Ab initio Simulation Package (VASP).[8-10] Perdew-Burke-Ernzerhof (PBE, which does not account the van der Waals (vdW) interaction),[11] PBE-D2,[12] or vdW-DF (optPBE-vdW, optB88-vdW, and optB86b-vdW)[13-14] functional is used to describe the exchange-correlation (XC) interaction depending on calculation type. Atomic relaxations are carried out until the internal forces are smaller than 0.01 eV/Å. The ab initio molecular dynamics (AIMD) simulations are performed within the canonical ensemble using the Nosé thermostat[15-16] with a time step of about 1-1.5 fs depending on the annealing temperature. The cutoff energy for the plane-wave basis is set to 300 eV for all AIMD simulations and 500 eV for all other calculations. To predict the energy convex hull, we form the dataset of competing phases using all structures available in Inorganic Crystal Structure Database (ICSD)[17] and Materials Project.[18] For each system, random atomic displacements with an amplitude of 0.1 Å are introduced to avoid trapping at a local energy minimum.

The phonon spectra are calculated using Phonopy code.[19] The obtained results are analyzed using Visualization for Electronic and Structural Analysis (VESTA) program[20] and pymatgen[21] library.

**Results and discussion**

**Energy filter for the stability of 2d materials:** From the thermodynamic perspective, the formation of any free-standing single 2d layer from its 3d counterpart requires energy (exfoliation energy) to break weak interlayer bonds holding the bulk structure together (see Fig. 1a). Learning from experiments, one can expect that only 2d materials within a specific range of exfoliation energies above a reference state can be synthesized as free-standing layers. In principle, this criterion allows using only 2d structures and corresponding bulk compounds to predict the materials stability. As an illustration, the energy to isolate single-layer graphene from bulk graphite is 0.07 eV/atom according to the optB86b-vdW results. However, this descriptor does not guarantee the materials stability with respect to other competing phases. Indeed, many layered compounds are not stable under ambient conditions. Moreover, it is not clear how the exfoliation energy can be calculated for single-layer materials without the bulk analogs, which is the case for a wide range of emerging 2d compounds. For instance, MXene compounds (*i.e.*, metal carbides and carbonitrides), metal oxides, and metal hydroxides are often obtained via controllable etching of selected atomic layers from the bulk compounds, which is followed by the mechanical exfoliation.[22] According to the basic thermodynamics, the competition between all different phases can be effectively described by the energy convex hull. By definition, the energy convex hull[23] defines phases with energies below those for all possible mixtures of competing phases with the same compositions. A simplified case of binary Zr-S is illustrated in Fig. 1b, where the phases above the convex hull have a tendency to decompose while the phases on the convex hull are the ground state compounds. The concept of the convex hull can be extended to a finite temperature by including temperature contribution to the Gibbs free energy.[19, 24] While the preliminary ground state convex hulls for many bulk systems are available[18, 25], to the best of our knowledge, only a few works[3, 26] applied this concept to 2d systems. Since synthesis by exfoliation implies breaking of interlayer bonds, which is an endothermic process, all 2d materials with 3d counterparts naturally fall above the convex hulls. As expected, 2d-ZrS$_2$(p-3m1) is found to be above the convex hull with the exfoliation energy of 0.09 eV/atom according to the optB86b-vdW results (in this work, the layer groups are used to describe symmetry of 2d materials utilizing Bilbao Crystallographic Server[27]). *Herein, for the energy filter, to analyze the 2d material stability in the thermodynamic limit, we use the isolation energy computed as the difference of formation energies of the target 2d material and convex hull at the specific composition (see Fig. 1). When the bulk counterpart of the target 2d material is the ground state compound, this energy is equivalent to the exfoliation energy as shown on the example of 2d-ZrS$_2$(p-3m1) in Fig. 1b. However, when 2d material does not have the bulk counterpart or the counterpart is above the convex hull, this energy represents the energy cost to form 2d material with respect to ground state compounds.* In principle, such descriptor allows to identify all 2d materials that can be synthesized when all competing phases are known. It should be noted, however, that this descriptor neglects kinetic factors; *i.e.*, one cannot predict if metastable phases can be realized due to slow kinetics of the phase transition. As an example, this criterion cannot describe the stability of fullerene with respect to graphite. Nevertheless, for simplicity, we do not include the kinetic factors into consideration.

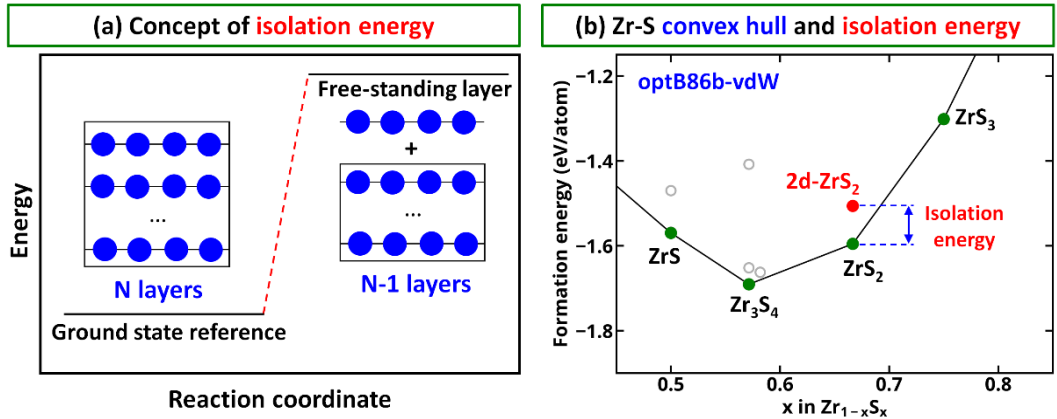

**Figure 1.** (a) Schematic illustration of the isolation energy defined as the energy needed to create a free-standing 2d layer from the bulk ground state reference. When target 2d material has bulk ground state counterpart, the isolation and exfoliation energies are equivalent to each other. Otherwise, the isolation energy represents the energy cost to form 2d material with respect to ground state compounds. (b) Energy convex hull of Zr-S system showing the amount of energy needed to isolate 2d-ZrS$_2$(p-3m1) from layered bulk ZrS$_2$(P-3m1). Green and non-filled gray markers represent ground state compounds and materials above the convex hull, respectively. Formation energy of the 2d material is shown as the red marker.

**Assembling a structure dataset:** To determine materials stability, it is critical to form an exhaustive dataset of all competing phases. Although preliminary theoretical results for convex hulls of bulk materials are available in Materials Project[18] and Open Quantum Materials Database (OQMD)[25], the materials dataset used in each database separately is incomplete. Every year new compounds are synthesized and predicted via different computational schemes.[28-30] Though it is only a matter of time until all competing phases are identified, *the usage of incomplete structure dataset existing so far creates two possible scenarios that can affect conclusions on stability of 2d materials: (I) unknown layered compounds can be the source for highly stable 2d materials; (II) important competing phases can be excluded from the convex hull analysis, which can result in predicting unstable ground state compounds and hence unstable 2d materials*. On the one hand, the scenario I opens the possibility of using different structure prediction approaches to find new ground state compounds and potential 2d materials. On the other hand, the scenario II has a strong deceptive aspect as the proposing unstable 2d materials for a specific application can mislead other scientists. To demonstrate scenario I, we consider the Si-S system. Specifically, we find that single layer 2d-SiS(pma2) structure[31] found from global optimization algorithm[32] is only about 0.13 eV/atom about the convex hull according to the optB86b-vdW calculations (see Figs. 2a,b), which is comparable to that for recently synthesized 2d materials (see discussion below). These results thus do not exclude the possibility to realize 2d-SiS(pma2) experimentally. However, since the 2d material does not have bulk counterparts, its potential realization will require development of new synthesis strategies. To demonstrate scenario II, we study the stability of 2d-PO(pm2$_1$n) material[33] found from O decoration of black phosphorene (see Figs. 2c), which also has recently received attention due to the high reactivity of phosphorene.[34-35] When only elemental reference states(i.e., bulk P and molecular O$_2$) are taken into account, one can conclude that due to negative formation energy, 2d-PO(pm2$_1$n) is stable. However, 2d-PO(pm2$_1$n) is 0.56 eV/atom above the convex hull according to the optB86b-vdW results when binary P-O bulk phases are included as shown in Fig. 2d. This energy is almost 4 times larger than that for any synthesized material discussed below. It should also be noted that a few recent works have shown that 2d-PO(pm2$_1$n) structure is high energy compound which can reconstruct to lower energy systems even within the same composition.[36-37] Because of these, we conclude that 2d-PO(pm2$_1$n) cannot be realized experimentally in free-standing single-layer form.

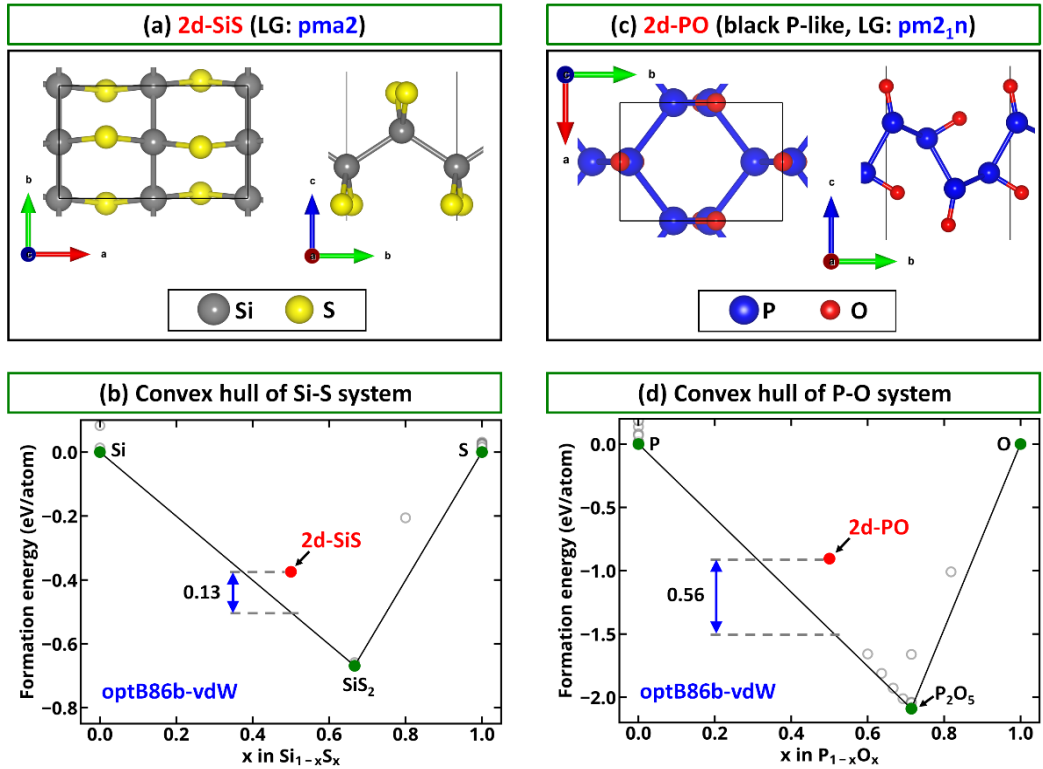

**Figure 2.** Examples of how incomplete structure dataset can affect conclusions concerning the discovery of new 2d materials. (a) 2d-SiS(pma2) structure which does not have experimentally observed bulk counterparts. (b) Si-S energy convex hull showing that 2d-SiS(pma2) is only 0.13 eV/atom above the convex hull. This energy is comparable to the isolation energies to form other known 2d materials. (c) 2d-PO(pm2$_1$n) found from the oxidation of blue phosphorene, the 2d material does not have experimentally observed bulk counterparts. (d) P-O energy convex hull showing that 2d-PO(pm2$_1$n) has negative formation energy but is 0.56 eV/atom above the convex hull when binary P-O competing phases are taken into account. This energy is about 4 times larger than the isolation energy for any known 2d material.

**Instability of experimentally known competing phases:** The majority of crystalline data available in different materials databases are experimentally observed (*e.g.*, taken directly from ICSD[17]) or based on direct first-principles optimizations of experimental structures with preserved crystal symmetries. Of course, since first-principles calculations involve a number of various approximations, one can argue that experimentally synthesized structures should be the ground states as they are "*real-life*" observations. It, however, appears that X-ray powder diffraction (XRD) refinement and its assignment to specific crystal structures are complex tasks.[38] For example, the same XRD patterns (especially for complex multicomponent systems) can be assigned to different crystal structures. The additional complexity arises from detecting materials composition, which is not the simplest thing to measure with high precision as many materials can exhibit non-stoichiometry.[39-43] Hence, *preserving high* *crystal symmetries of experimentally observed phases during first-principles calculations can be another limitation in building the convex hull.* In fact, standard experimental characterization techniques provide the averaged atomic positions over the characterization time.[44] As a consequence, for some materials, high symmetry positions reported in experimental studies can be an artifact – atoms may prefer to form different local environments[45] or have noticeable structural changes to minimize the Gibbs free energy.

**Role of the XC functional:** To study the stability of 2d materials, the vdW forces should be described accurately. The main problem originates from the limitation of classical XC functionals (*e.g.*, PBE[11] or PBEsol[46] within the generalized gradient approximation) to describe the long-range interaction. Interlayer interaction in graphite is the simplest illustration of such limitation. Specifically, for graphite, PBE calculations provide unphysically large separation and minuscule interlayer interaction energy.[47] To overcome the limitations, various approaches to account the vdW interaction (*e.g.*, vdW-DF[48] and DFT-D[49]) have been developed. Despite this, recent studies have shown that the prediction of vdW forces remains the challenge despite new developments in the vdW field.[50-51] Specifically, it has been shown that all common approaches to treat vdW forces overestimate interlayer interaction energy in layered systems when compared to diffusion Monte-Carlo simulations.[50] Moreover, existing XC functional has a limitation to describe chemical potentials of elemental reference states.[52-54] Considering the diversity of existing XC functionals and their accuracy, it is not possible to determine a universal threshold isolation energy below which 2d materials can be synthesized. Moreover, using different XC functionals results in the difference in energy convex hulls as shown on the example of In-Se system in Fig. 3a,b. Specifically, for optB86b-vdW functional, the chemical potential stability range of InSe compound is defined by competing with $In_4Se_3$ and $In_6Se_7$ phases; while for PBE-D2, it is determined by the competition with In and $In_2Se_3$ compounds instead. This discrepancy in the results is due to the $In_4Se_3$ and $In_6Se_7$ phases being 39 and 3 meV/atom above the convex hull in PBE-D2 calculations. *Despite the difference in convex hulls for different functionals, we believe that one can define threshold isolation energy for each XC functional as the maximum isolation energy computed for already synthesized 2d materials.* Because of this, based on detailed literature analysis, we form a list of known free-standing 2d materials. It should be noted that we exclude 2d materials which were synthesized on substrates as the substrates play an important role in stabilization of 2d materials as known on an example of silicene at metal surfaces.[7] In Fig. 3c, we summarize the results for isolation energies of 2d materials obtained with different XC functionals. To our surprise, despite a fundamental difference in approaches how DFT-D and vdW-DF functionals treat the vdW forces, the expected threshold isolation energies computed for different XC functionals are around 0.11-0.12 eV/atom. The PBE calculations provide the lowest threshold isolation energy of 0.04 eV/atom among all considered functionals (one should note however that the PBE convex hull calculations can have significant error bar due to neglecting the vdW forces which play important role in layered materials). These results suggest that threshold isolation energies should be adjusted to specific XC functional especially when the functional cannot capture the vdW forces. In the above discussion, we do not take into consideration 2d-Te phases as they have isolation energies of about 2 times larger (0.21 eV/atom according to the optB86b-vdW calculations) than that for any other materials. These results might be due to inaccuracy of the XC functionals, but it is also possible that 2d-Te realized experimentally was stabilized by other factors, such as surface passivation, multilayer thickness, or substrate effects.

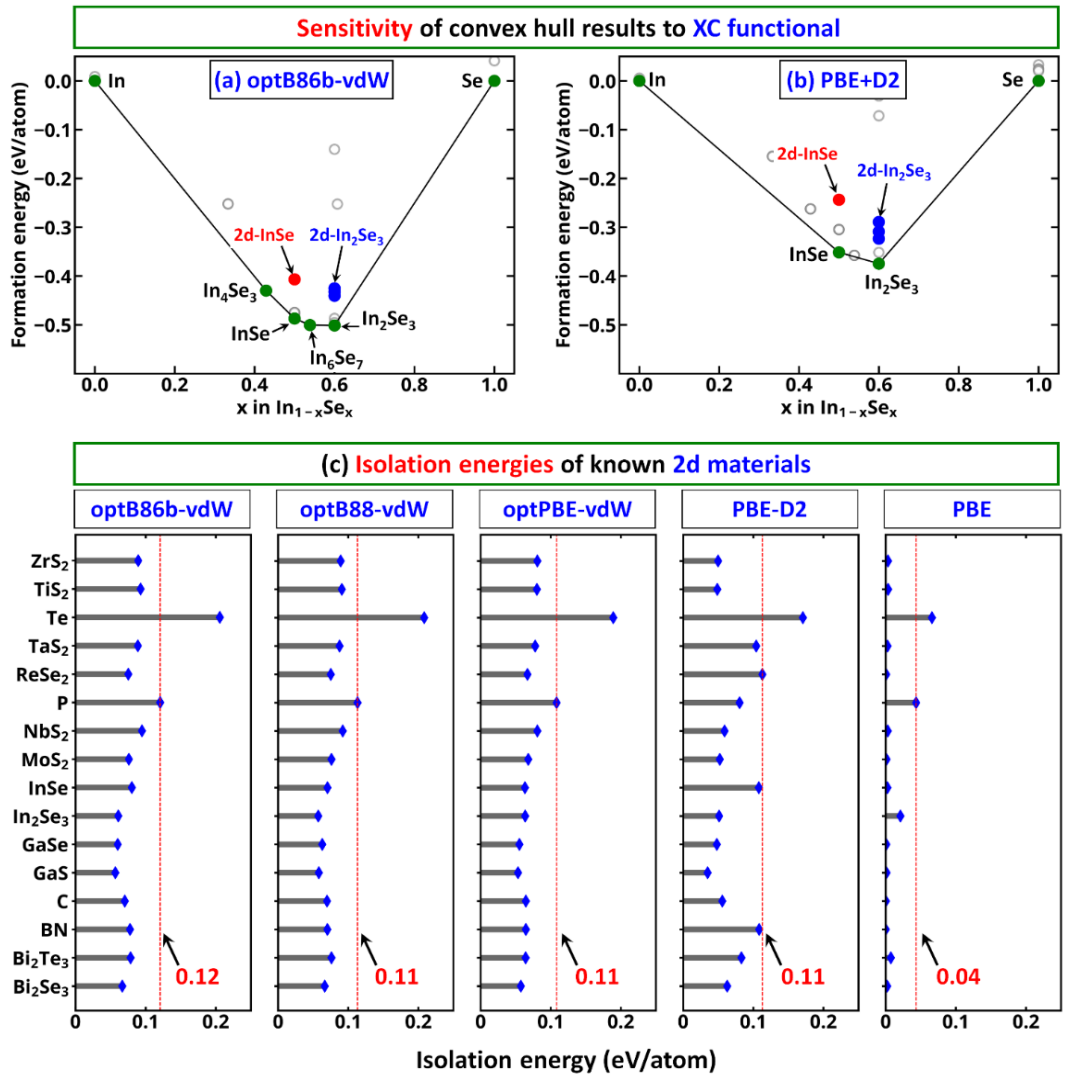

**Figure 3.** Role of the XC functional in the calculations of energy convex hulls and isolation energies. In-Se convex hulls computed using (a) optB86b-vdW and (b) PBE-D2 functionals. Formation energies of 2d-materials are shown as red and blue markers. (c) Variation of isolation energies for 16 materials synthesized experimentally for different XC functionals. The expected threshold isolation energy computed for each XC functional is shown by red dashed line.

**Phonon filter for the stability of 2d materials:** *At equilibrium, the potential energy of the system always increases with respect to any combinations of atomic movements. This allows to use vibrational spectra as the filter to validate materials stability as the presence of imaginary frequency signifies material instability.* To demonstrate the effectiveness of phonon filter, we consider 2d-$MoSe_2$(p-6m2) and 2d-$NbS_2$(p-3m1) (see Fig. 4a,c) isolated from bulk $MoSe_2$(P-6m2) and $NbS_2$(P-3m1), respectively. For 2d-$MoSe_2$(p-6m2), the vibrational band structure does not have regions with the imaginary frequency which imply that the phonon stability of the 2d material (see Fig. 4b). The case is however very different for 2d-$NbS_2$(p-3m1) where the existence of imaginary frequencies is observed as shown in Fig. 4d. These results suggest that 2d-$NbS_2$(p-3m1) is unlikely to be realized experimentally. It should be noted that the phonon filter is the most widely used in the stability analysis of 2d materials and a large number of phonon spectra has recently become publicly available within Materials Cloud database.[55] Despite such popularity, we regard satisfying the phonon filter as necessary but not sufficient condition to evince dynamic stability of the material. The reason for such conclusion is that the phonon analysis deals with small atomic displacements, and therefore, is unable to capture phase transitions coupled with complex lattice reconstructions.

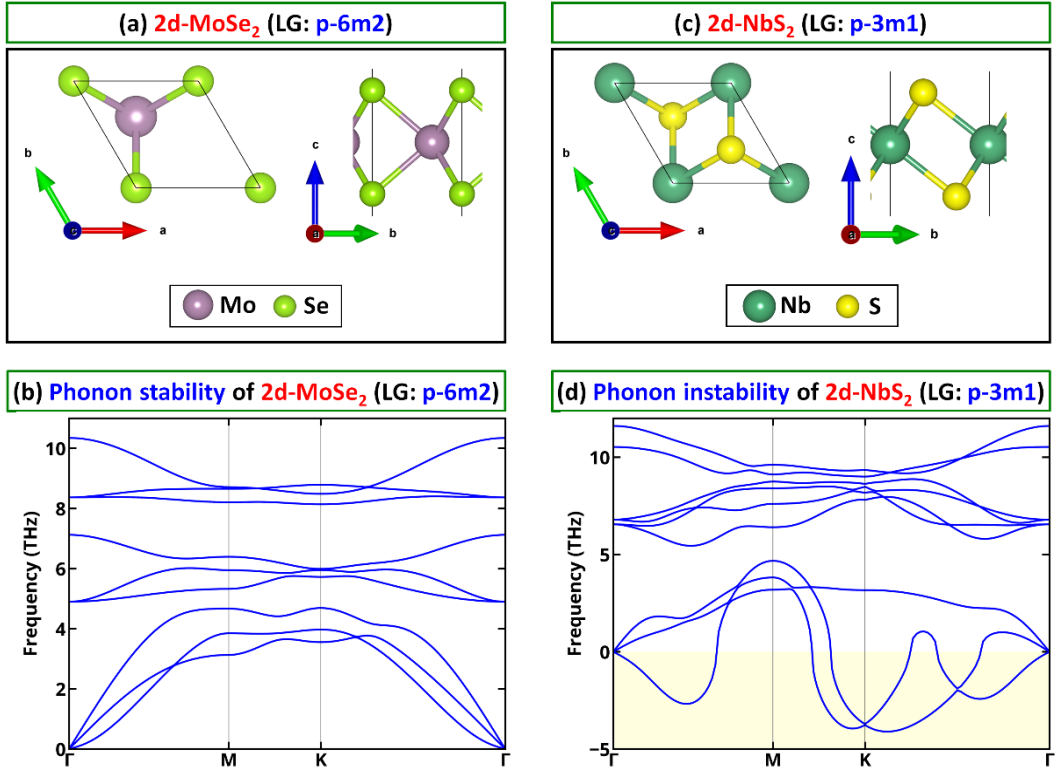

**Figure 4.** Examples of application of phonon filter to analyze the stability of 2d materials. (a) 2d-MoSe$_2$(p-6m2) and (c) 2d-NbS$_2$(p-3m1) structures isolated from MoSe$_2$(P-6m2) and NbS$_2$(P-3m1) bulk structures, respectively. Phonon dispersions for (b) 2d-MoSe$_2$(p-6m2) and (d) 2d-NbS$_2$(p-3m1). The results for 2d-MoSe$_2$(p-6m2) show that the system does not have imaginary vibrational frequencies, while for 2d-NbS$_2$(p-3m1), the results suggest that the system cannot be realized in the free-standing form due to the presence of imaginary frequencies (shown as negative frequencies).

**Dynamic filter for the stability of 2d materials:** To ensure that a material is dynamically stable, one need to evince that it cannot undergo any structural changes that lower its energy. In practice, at the first-principles level, the verification can be done using AIMD simulations. As an illustration, we consider recently proposed 2d-SiP(p3m1)[56] found from a partial substitution of Si atoms in silicene by P and 2d-PO(p-3m1)[57] proposed based on O decoration of blue phosphorene (see Fig. 5a,d). It should be noted that 2d-PO(p-3m1) has no imaginary frequencies in its vibrational spectra.[57] To verify the dynamic stability of the proposed 2d materials, we employ AIMD simulations with one annealing step at a fixed temperature (the temperature protocols can be different and include for instance quenching steps or similar). Every 9 ps, we extract the structure and perform direct structure relaxation without any quenching protocol. This step is mainly done to verify the effect of atomic motions on the system's energy excluding temperature contribution. Indeed, *since the size of modeled systems in AIMD simulations is limited to a few hundreds of atoms, the energy fluctuation can be larger than energy lowering due to structural changes. Hence, one cannot use the potential energy profile only as the criterion for the dynamic stability of the materials.* The computed energy profiles for 2d-SiP(p3m1) and 2d-PO(p-3m1) show that both potential and formation energies exhibit stair-like behavior where the structural changes result in the formation of lower energy structures. As a consequence, the 2d-SiP(p3m1) and 2d-PO(p-3m1) are not likely to be realized experimentally in the free-standing forms. One should note however that these results do not indicate that other 2d-SiP and 2d-PO structures cannot be synthesized. As an example, 2d-SiP(c2/m11) isolated from SiP(Cmc2$_1$) has high dynamic stability and negative formation energy.[51]

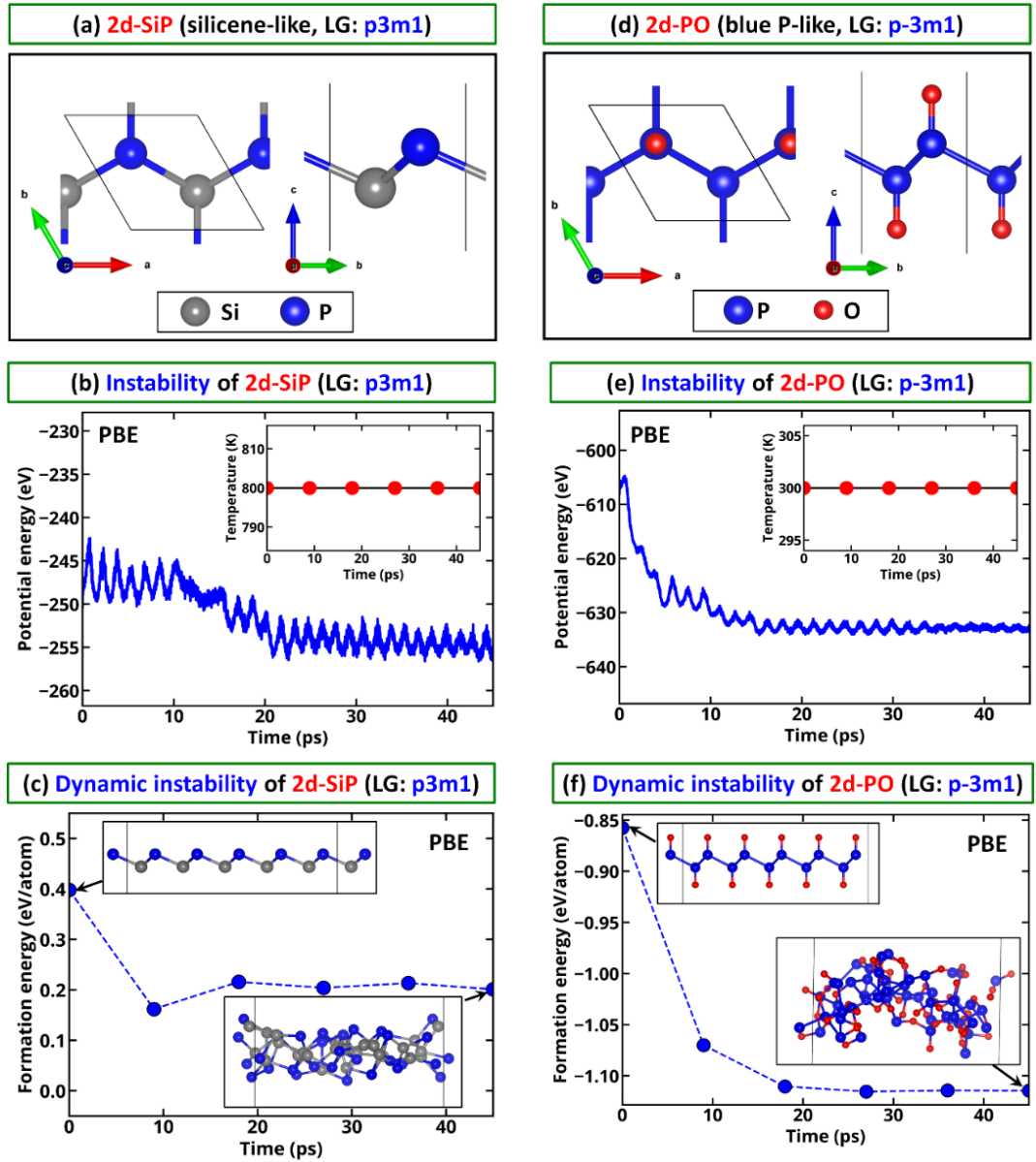

**Figure 5.** Examples of application of the dynamic filter to analyze the stability of 2d materials. (a) 2d-SiP(p3m1) structure found based on partial replacement of Si by P in silicene. (d) 2d-PO(p-3m1) structure obtained by oxidation of blue phosphorene. (b,e) Potential and (c,f) formation energy profiles for 2d-SiP(p3m1) and 2d-PO(p-3m1) structures during the AIMD simulations. The results show that both 2d-SiP(p3m1) and 2d-PO(p-3m1) tend to reconstruct to lower energy structures and are unlikely to be realized experimentally in free-standing forms. The insets in (b,e) show the schematic illustrations of temperature protocols used in the AIMD simulations.

The stair-like behavior for potential energy profiles observed for both for 2d-SiP(p3m1) and 2d-PO(p-3m1) can be understood from the diffusion perspective as structural changes requires atomic diffusion where the frequency of diffusion jumps ($\Gamma$) is defined by the Arrhenius law ($\Gamma \sim \exp(-\Delta E/kT)$, where $\Delta E$, k, and T are diffusion barrier, Boltzmann constant, and temperature). Considering that the typical time for a diffusion jump is in order of ps, usually, it is not possible to reach global energy minimum starting from any random structures as time-scale even for state-of-art AIMD simulations is limited to a few hundreds of ps. Moreover, the frequency of diffusion jumps is defined by the diffusion barrier and temperature. Because of this, *a few ps AIMD simulation at room temperature is likely to be insufficient for predicting dynamic stability even if the potential energy and formation energy profiles do not change with time*. To demonstrate this limitation, we use recently proposed 2d-SiS(pm2$_1$n) structures (see Fig. 6a) inspired by black phosphorene which is also known to have no imaginary vibrational frequencies.[58] The energy profiles for 2d-SiS(pm2$_1$n) at 300 K show that the system does not have any structural changes during 45 ps (Fig. 6b). This behavior is consistent with the analysis of formation energies of extracted systems that remain roughly the constant during the AIMD simulation (Fig. 6d). It does not guarantee however that the system is dynamically stable. Specifically, the potential energy and formation energy profiles computed for AIMD simulation at 600 K show that system has the tendency to reconstruct in lower energy configurations (see Fig. 6c,d). These results highlight the necessity of a detailed description of the dynamic evolution of proposed 2d materials over a wide range of temperatures and time to avoid potentially misleading conclusions on materials stability.

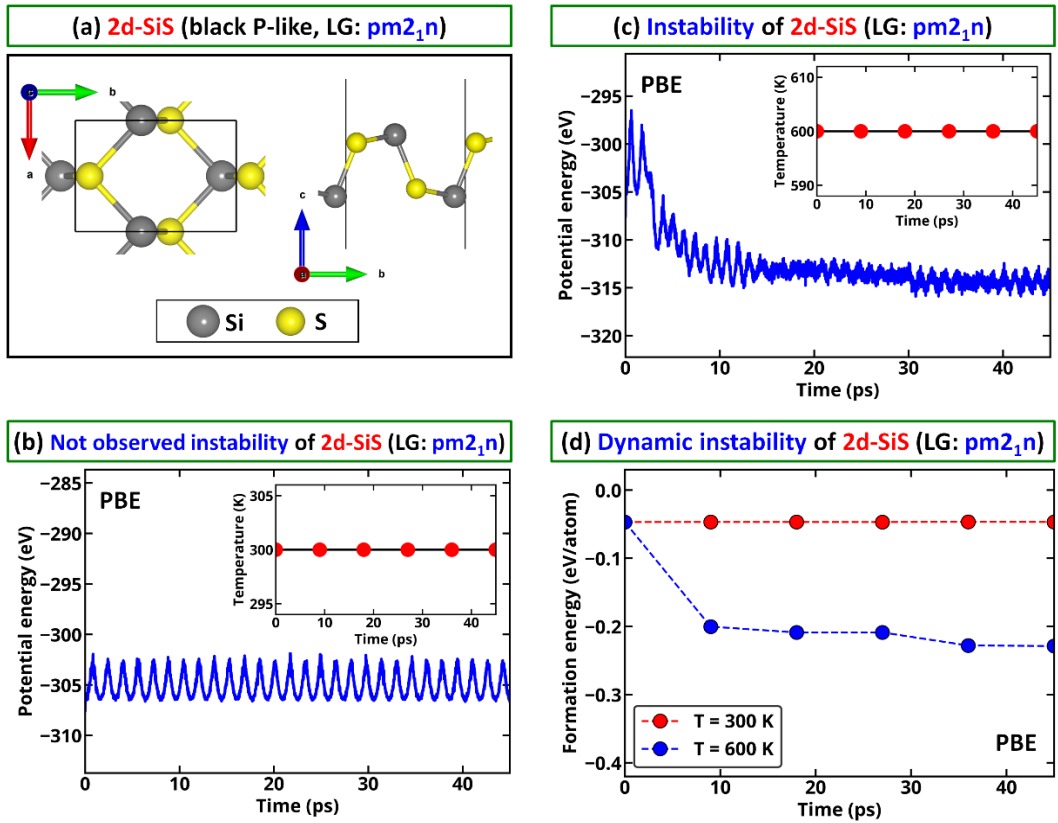

**Figure 6.** Role of temperature protocol in the analysis of dynamic stability of 2d materials. (a) 2d-SiS(pm2$_1$n) structure found based on substitution of P atoms in black phosphorene by Si and S atoms. Potential energy profiles for 2d-SiS(pm2$_1$n) during the AIMD simulation at (b) 300 K and (c) 600 K. The insets in (b) and (c) schematically illustrate the temperature protocols used in the AIMD simulations. (d) Formation energies of structures taken from the AIMD simulation at different moments of time.

**Conclusions:** Considering a number of already synthesized and recently proposed 2d materials, we construct energy, phonon, and dynamic filters to study the stability of potential 2d compounds.

Specifically, by analyzing energy convex hull for already synthesized 2d materials, we propose the threshold isolation energy *(energy filter)* for each functional below which free-standing single-layer 2d materials can be isolated. This threshold isolation energy is sensitive to XC functional used in the calculations and need to be adjusted for specific computational setup. The computed threshold isolation energies are 0.12, 0.11, 0.11, 0.11, and 0.04 eV/atom for optB86b-vdW, optB88-vdW, optPBE-vdW, PBE-D2, and PBE functionals, respectively. Based on energy filter, we conclude that one can probably not realize 2d-PO($pm2_1n$) due to the high energy cost to form it with respect to other competing phases (the system has the isolation energy about 4 times larger than the threshold isolation energy according to the optB86b-vdW calculations). Taking into account that at equilibrium the potential energy of the system always increases with respect to any combinations of atomic movements, we demonstrate the effectiveness of the analysis of vibrational spectra *(phonon filter)* on the example of $NbS_2$(P-3m1) that is unlikely to be realized due to imaginary vibrational frequencies in its vibrational spectra. We also explain that satisfying the phonon filter is a necessary but still insufficient condition to analyze dynamic stability of the target material. In particular, we show by AIMD simulations *(dynamic filter)* that 2d-PiO(p-3m1) and 2d-SiS($pm2_1n$) structures which both do not have imaginary vibrational frequencies can still be dynamically unstable. Moreover, our results highlight the necessity of a detailed description of the dynamic evolution of proposed 2d materials over a wide range of temperatures and time to avoid potentially misleading conclusions on the dynamic stability of 2d materials. We believe that application of the above-constructed filters for the search of novel 2d materials can reduce the gap between the theoretical predictions of novel 2d materials and their experimental realization.

**Acknowledgments:** This work is financially supported by the Research Council of Norway (ToppForsk project: 251131). We acknowledge Norwegian Metacenter for Computational Science (NOTUR) and Swedish National Infrastructure for Computing (SNIC) for providing access to supercomputer resources.